
\magnification=\magstep1
\parskip 0pt
\parindent 15pt
\baselineskip 16pt
\hsize 5.53 truein
\vsize 8.5 truein

\font\titolo = cmbx9 scaled \magstep2
\font\autori = cmsl9 scaled \magstep2
\font\address = cmr8
\font\abstract = cmr9

\hrule height 0pt
\rightline{POLFIS-TH.03/93}
\vskip 1truein
\centerline{\titolo LOW--TEMPERATURE EXPANSION FOR A}
\vskip .05truein
\centerline{\titolo FIRST ORDER SURFACE TRANSITION}
\vskip .15truein
\centerline{\autori Carla Buzano and Alessandro Pelizzola}
\vskip .15truein
\centerline{\address Dipartimento di Fisica and Unit\'a INFM,
Politecnico di Torino, I-10129 Torino, Italy}

\vfill

\centerline{\bf ABSTRACT}
\vskip .05 truein {\abstract The question concerning the possibility of a
first order surface transition in a semi--infinite Blume--Capel model is
addressed by means of low temperature expansions. It is found that such a
transition can exist, according to mean field and cluster variation
approximations, and contrarily to renormalization group results. }
\vskip .35 truein
\vfill
\eject

\parindent 15pt

The semi--infinite spin--1 Blume--Capel model
has been recently introduced to describe the surface critical behavior of
magnetic systems[1] and $^3$He--$^4$He mixtures[2]. The phase diagram of
this model has been investigated by means of some well--known methods,
namely mean field approximation[1--3] (MFA), real space renormalization
group[4,5] (RG) and cluster variation method[6] (CVM). All these methods
agree to show that the model can exhibit three different phases: a
completely disordered phase (D), a completely ordered phase (O) and a
partially ordered phase (S), in which order is localized on the surface of
the lattice. Only very recently it has been recognized[3] that, at least at the
mean field level, for certain values of the model parameters, two different O
phases can be found, with different degrees of order at the surface. \par
Accordingly, different species of phase transitions are found. Lowering the
temperature the model can either order as a whole, passing from the D to the O
phase, and the transition is said to be {\it ordinary}, or it can undergo
two different transitions, a {\it surface} one from the D to the S phase
and then, at a lower temperature, an {\it extraordinary} one from the S to
the O phase. In the limiting case between the two above, the transition is
said to be {\it special}. \par
In the MFA phase diagram[1] all these transitions can be either second or
first order, depending on the model parameters (only the transition between
the two O phases considered in [3] is always first order),
and this is confirmed by
the CVM analysis[6]. On the other hand, the RG scheme developed in [4] (but
not the one in [5])
excludes the possibility of extraordinary or surface first order
transitions at finite (i.e. nonzero) temperatures. It is then legitimate to ask
whether such transitions do really occur or they are only features
introduced by some particular approximation. \par
In the present Letter we have addressed this question by means of low
temperature expansions. By using a technique similar to that outlined in
[7] in the context of the Pirogov--Sinai theory (although nothing is
rigorous for semi--infinite systems) we show that a surface first order
phase transition can be found at low, but nonzero, temperature, and
this also implies the existence of an extraordinary first order transition.
\par
Our method consists in expanding the free energies of the S and D phases
taking into account only a few excitations over the corresponding ground
states. The phase boundary (a first order transition line) is then
determined by comparison of these free energy. \par
Let us consider the Blume--Capel model on a semi--infinite lattice with a
free surface, which is described by the following hamiltonian:
$$- \beta H = J_s \sum_{\langle i j \rangle} S_i S_j - \Delta_s \sum_i S_i^2
+ J_b \sum_{\langle k l \rangle} S_k S_l - \Delta_b \sum_k S_k^2,\eqno(1)$$
where $S_i = +1, 0 , -1$, $\displaystyle{\sum_{\langle i j \rangle}}$
denotes a sum over all nearest neighbors (n.n.) with both sites lying on
the surface, $\displaystyle{\sum_{\langle k l \rangle}}$ denotes a sum over
the remaining n.n., and $\beta = (k_B T)^{-1}$ (with $k_B$ Boltzmann
constant and $T$ absolute temperature). $J_s$ and $J_b$ (both positive,
since we limit ourselves to the ferromagnetic case) are reduced surface and
bulk exchange interactions, while $\Delta_s$ and $\Delta_b$ are reduced
surface and bulk anisotropy, respectively. \par
In the following we will
treat the case $J_b = J_s = J$, $D = \Delta_b/\Delta_s > 3/2$ on a simple
cubic lattice with a (100) free surface.
In such a case (see e.g. [3]) the ground state is O for $\Delta_b < 3 J$,
S for $3 J < \Delta_b < 2 D J$ and D for $\Delta_b > 2 D J$ and, according
to mean field theory, in the $(\Delta_b/6J,T)$ phase diagram, first order
extraordinary and surface transition lines start from $(1/2,0)$ and
$(D/3,0)$ respectively. Let us now concentrate on the surface transition
and develop low temperature expansions for the surface contributions
$f_s^{(S)}$ and $f_s^{(D)}$ to the free energies of the S and D phases. \par
To begin with, let us consider the S phase, which is characterized by $S_i
= +1$ (or, equivalently, $-1$) for all surface sites, and $S_k = 0$ for all
bulk sites. On a lattice with $N$ surface sites, its reduced ground state
energy is
$E_0^{(S)} = N(-2J + \Delta_s)$. To determine a low temperature expansion
for $f_s^{(S)}$ up to the 5th order in $x = e^{-J}$ we have enumerated all
the excitations with respect to the ground state with energy $\delta E < 6
J$. An excitation is regarded
as a set of spins which are in a state different from the
ground state, and is represented by a graph, obtained drawing a set of circles
containing the values
of the changed spins and joining with a line each pair of neighboring spins.
We have then surface, bulk and mixed excitations, depending on which spins
are changed. To each excitation $g$ (see Tab. I for some examples) we have
assigned a multiplicity $M(g)$, which
is the number of embeddings of the corresponding graph in the lattice
under consideration and a Boltzmann weight $e^{-\delta E(g)}$. \par
We can thus write the partition function in the form
$$Z^{(S)} = e^{-E_0^{(S)}}\sum_g M(g) e^{-\delta E(g)}, \eqno(2)$$
which, up to the 5th order in $x$, reads (for a lattice of L layers, each
of N sites)
$$\eqalign{Z^{(S)}&(N,L) = \cr
& 2 w^{-N} \Biggl\{ 1 + N \Biggl[ (w + y)x^2
+ 2(w^2 + (L - 2)y + y^2)x^3 \cr
& + \Biggl( {N-5 \over 2}w^2 + 6w^3 + w^4 + y + {N-3 \over 2}y^2
+ 6y^3 + y^4 + (N-1)w y \Biggr)x^4 \cr
& +2 \Bigl( (N-8)w^3 + 9w^4 + 4w^5 + w^6 \cr
& + (NL - 2N + 3L - 8)y^2 + (N-6)y^3 + 9y^4 + 4y^5 + y^6 \cr
&    + (N(L-2)+1)w y + (N-2)w^2 y +
(N-2)w y^2 \Bigr) x^5 \Biggr] \Biggr\}, \cr} \eqno(3)$$
where $w = e^{\Delta_s - 2 J}$, $y = w^{-D}x^{2\epsilon}$ and
$\epsilon = D - 3/2$. \par
Given a partition function $Z(N,L)$, the surface free energy density of the
corresponding semi--infinite system is given by [8]
$$f_s = \lim_{N,L \to \infty} {- \ln Z(N,L) - NL f_b \over N},\eqno(4)$$
where $f_b$ is the bulk free energy density and is given by
$$f_b = -\lim_{N,L \to \infty} {\ln Z(N,L) \over NL}. \eqno(5)$$
We then have, for the surface free energy density of the S phase:
$$\eqalign{f_s^{(S)} = & \ \Delta_s - 2 J - \Biggl\{ (w + y)x^2 +
2(w^2 - 2y + y^2)x^3  \cr
& + \Biggl[ w^2 \Biggl( w^2 + 6w - {5 \over 2} \Biggr) + (1-w)y
- {3 \over 2} y^2+ 6y^3 + y^4 \Biggr] x^4 \cr
& + 2 \Bigl[ w^3(w^3 + 4w^2 + 9w - 8) \cr
& + (1 - 2w)wy - 2(w+4)y^2
- 6y^3 + 9y^4 + 4y^5 + y^6
\Bigr] x^5 \Biggr\}. \cr}\eqno(6)$$ \par
In the same way we find
$$\eqalign{Z^{(D)}&(N,L) = \cr
& 1 + 2N\Bigl\{ w^{-1}x^2 + \Bigl[ 2w^{-2} +
(L-1)y \Bigr] x^3  \cr
& + \Bigl[ w^{-2}(w^{-2} + 6w^{-1} + N-5) + w^{-1}y \Bigr] x^4 \cr
& + \Bigl[ 2w^{-2} \Bigl( w^{-4} + 4w^{-3} + 9w^{-2} + 2(N-8)w^{-1} + 1
\Bigr)  \cr
& + 2(NL - N - 1)w^{-1}y + 4w^{-2}y
+ (3L-4)y^2 \Bigr] x^5 \Bigr\} \cr} \eqno(7)$$
and
$$
\eqalign{f_s^{(D)} = & -2\Bigl\{ w^{-1}x^2 + (2w^{-2} - y)x^3
 \cr
& + \Bigl[ w^{-2}(w^{-2} + 6w^{-1} - 5) + w^{-1}y \Bigr] x^4 \cr
& + \Bigl[ 2w^{-2}(w^{-4} + 4w^{-3} + 9w^{-2} - 16w^{-1} + 1)
 \cr
&   -
2(1-2w^{-1})w^{-1}y - 4y^2 \Bigr] x^5 \Bigr\} . \cr} \eqno(8)$$ \par
The transition line is then readily obtained by numerically solving the
equation $f_s^{(S)} = f_s^{(D)}$ for $\Delta_b/6J$ at each temperature, and
is reported in Fig. 1 together with MFA and CVM results, for a choice of
values of the model parameters. The transition is
first order, as can be checked by looking at derivatives of the free
energies, and the results of MFA and CVM are qualitatively confirmed (the
reentrance predicted by CVM is not seen here).
The above results should be particularly reliable for $\epsilon < 1/12$ (in
order to satisfy $\delta E(g) < 6 J$ for all the excitations considered) and
for $k_B T /6 J < 0.06$, so that the 5th order
contributions are no more than a few percent of the sum of the lower order
terms (excluding the 0th order ones, of course, since they are defined up to an
additive constant), and these conditions are satisfied in Fig. 1. \par
Implicitly, since the bulk phase diagram is simply the phase diagram of the
infinite system, which is known to undergo also a first order transition,
our results show that the Blume--Capel model on a semi--infinite lattice is
also capable of exhibiting an extraordinary first order transition. \par
Summarizing, we have shown by low temperature expansions that the surface
transition of the semi--infinite Blume--Capel model can be also first
order, according to previous results obtained by MFA and CVM, and in
contrast with a RG analysis. We believe that the present method is more
accurate than the RG scheme proposed in [4], since this kind of low
temperature expansion, even if not rigorous in the presence of a free
surface, is particularly well--suited for first order transitions.
Furthermore, RG schemes for models with several different phases and rich
phase diagrams heavily depend on the choice of the mapping truncation, as
can be seen by comparing the results in [4], which exclude the possibility
of a surface or extraordinary first order transition, with those in [5],
where, by means of a different RG scheme, such transitions are found.
Indeed, only very recently [9] Berker and Netz have succeeded in devising a
RG scheme capable of describing the antiquadrupolar and ferrimagnetic phases
of the Blume--Emery--Griffiths model. \par
\vfill\eject
{\bf References} \bigskip
\item{[1]} A. Benyoussef, N. Boccara and M. Saber, J. Phys. C: Solid State
Phys. {\bf 19} (1986) 1983.
\item{[2]} X.P. Yiang and M.R. Giri, J. Phys. C: Solid State Phys. {\bf 21}
(1988) 995.
\item{[3]} C. Buzano and A. Pelizzola, submitted to Phys. Rev. Lett.
\item{[4]} A. Benyoussef, N. Boccara and M. El Bouziani, Phys. Rev. B {\bf
34} (1986) 7775.
\item{[5]} L. Peliti and S. Leibler, J. Physique Lett. {\bf 45} (1984) L-591.
\item{[6]} C. Buzano and A. Pelizzola, Physica A {\bf 195} (1993) 197.
\item{[7]} J. Slawny, in {\it Phase Transitions and Critical Phenomena},
vol. 11, ed. by C. Domb and J.L. Lebowitz (Academic Press, London, 1987),
Chap. 3.
\item{[8]} R. Pandit, M. Schick and M. Wortis, Phys. Rev. B {\bf 26} (1982)
5112.
\item{[9]} A.N. Berker and R.R. Netz, Phys. Rev. B, in press.
\vfill\eject
{\bf Figure Caption}
\parindent .4 truein
\item{}
\itemitem{\hbox to .67 truein{Fig. 1: \hfill}} Surface first order
transition lines for $\epsilon = 0.03$ as given by the present method
(solid line), MFA (dashed line) and CVM (dotted line).
\vfill\eject

\parindent 0pt
\font\tenrm = cmr10
\font\forcirc = cmsy10 scaled \magstep1
\def\spinplus{\hbox{\forcirc \char'15 \kern-0.83em \tenrm + \kern-0.2em}}
\def\spinzero{\hbox{\forcirc \char'15 \kern-0.72em \tenrm 0 \kern+0.0em}}
\def\hbond{\hbox{\vrule width1.2em height2.4pt depth-2.0pt}}
\def\vbond{\hbox{\vrule width0.4pt height1.2em depth0.0pt}}

\def\surfexc{\vbox to 6em{\vfill
\spinzero
\hbox{\kern 0.56em \vbond} \par
\hbox {\hfill \spinzero \hbond \spinzero \hfill} \par
\hbox{\kern 0.56em \vbond} \par
\spinzero
\vfill}}

\def\bulkexc{\vbox to 6em{\vfill
\hbox {\vrule width 4.48em height 2.4pt depth -2.0pt
\tenrm Surface \hfill} \par
\vskip -0.2em
\hbox {\kern 1em \vbond \kern 2.28em \vbond \hfill} \par
\hbox {\kern .45em \spinplus \hbond \spinplus \hfill} \par
\hbox {\kern 1em \vbond \hfill} \par
\hbox {\kern .45em \spinplus \hfill}
\vfill}}

\def\mixexc{\vbox to 6em{\vfill
\hbox {\hbond \spinzero \vrule width 6.48em height 2.4pt depth -2.0pt
\tenrm Surface \hfill} \par
\vskip -0.5em
\hbox {\kern 5em \vbond \kern 2.28em \vbond \hfill} \par
\hbox {\kern 4.45em \spinplus \hbond \spinplus \hfill}
\vfill}}

\def\vcenter#1{\vbox to 6em{\vfill \hbox{#1} \vfill}}

{\bf Table I.} Examples of elementary excitations of the S phase \par
\bigskip
\vbox{\offinterlineskip
\hrule
\halign{&\vrule#&
  \strut\quad\hfil#\hfil\quad\cr
height2pt&\omit&&\omit&&\omit&&\omit&\cr
&$g$&&Type&&$M(g)$&&$\exp [-\delta E(g)]$&\cr
height2pt&\omit&&\omit&&\omit&&\omit&\cr
\noalign{\hrule}
height2pt&\omit&&\omit&&\omit&&\omit&\cr
&\surfexc&&\vcenter{Surface}&&\vcenter{$4N$}&&\vcenter{$w^4 x^5$}&\cr
&\bulkexc&&\vcenter{Bulk}&&\vcenter{$4N$}&&\vcenter{$y^3 x^5$}&\cr
&\mixexc&&\vcenter{Mixed}&&\vcenter{$2N(N-2)$}&&\vcenter{$w y^2 x^5$}&\cr
height2pt&\omit&&\omit&&\omit&&\omit&\cr}
\hrule}
\vfill \eject
\end